\documentclass[12pt]{article} \textheight=24 true cm
\usepackage{graphicx}
\usepackage{amsmath}
\usepackage{amssymb}
\usepackage{cite}
\usepackage{cleveref}
\usepackage{color}
\begin{document}
\begin{center}
{\Large\bf Thermodynamical behaviour of the Variable  Chaplygin gas }\\[8 mm]
D. Panigrahi\footnote{ Sree Chaitanya College, Habra 743268, India
\emph{and also} Relativity and Cosmology Research Center, Jadavpur
University, Kolkata - 700032, India , e-mail:
dibyendupanigrahi@yahoo.co.in }
   \\[10mm]

\end{center}
\begin{abstract}
The thermodynamical behaviour of the Variable Chaplygin gas (VCG)
model is studied, using an equation of state like $P = -
\frac{B}{\rho }$, where $B = B_{0}V^{-\frac{n}{3}}$. Here $B_{0}$ is
a positive universal constant,  $n$ is also a constant and $V$ is the volume of the fluid. From the
consideration of thermodynamic stability,
 it is seen that only if the  values of  $n$ are allowed to be negative, then  $ \left( \frac{\partial P}{\partial
 V}\right)_{S} <0$ throughout the evolution.
  Again thermal capacity at constant volume $c_{V}$ shows  positive expression.
 Using the best fit value of $ n = -3.4$ as previously found by Guo \emph{et al}~
 \cite{guo1} gives that the fluid is thermodynamically stable through out the evolution.
 The effective equation
 of state for the special case of, $n = 0$ goes to $\Lambda$CDM model.
 Again for $n <0$ it favours phantom-like cosmology which is in agreement with the
 current SNe Ia constraints like
 VCG model. The deceleration parameter is also studied in the context of
 thermodynamics and the  analysis shows that
 the  \emph{flip} occurs for the value of $ n < 4$. Finally the  thermal
 equation of state is discussed which is an explicit function of temperature
 only. It is also observed that the third law of thermodynamics is
 satisfied in this model.
 As expected the volume increases as temperature falls during adiabatic
expansions. In this case, for $ T \rightarrow 0$, the thermal
equation of state reduces to  $\left( - 1 + \frac{n}{6} \right)$
which is identical with the equation of state for the case of large
volume.

\end{abstract}
   ~~~~~~~KEYWORDS : cosmology;  chaplygin
   gas; thermodynamics

\hspace{0.2 cm} PACS :  05.70.Ce;98.80.Es;98.80.-k

\section*{1. Introduction}
\vspace{0.1 cm}
 Recent observational evidences suggest that
the present universe is accelerating ~\cite{reis, aman}. A Chaplygin
type of gas cosmology~\cite{kamen} is one of the plausible
explanations of recent phenomena, which  is a new matter field to
simulate dark energy. This type of equation of state(EOS) is not applicable in the case of primordial universe.
 This was discussed in the several articles
~\cite{fab, bena, lima, as, dp2}. Such equation of state leads to a component which
behaves as dust at early stage and as cosmological constant ($\Lambda$) at later stage. The form of the equation of state
(EoS) of matter is the following,

\begin{equation}
P = - \frac{B}{\rho }
\end{equation}
Here $P$ corresponds to the pressure of the fluid and $\rho$ is the
energy density of that fluid and $B$ is a constant. Recently a
variable Chaplygin gas (VCG) model was proposed and
constrained using SNeIa gold data~\cite{guo1,guo2} , assuming the
 $B$ to depend on scale factor of our metric chosen. Now we have taken the
above relation as  $B = B_{0}V^{-\frac{n}{3}}$ where  $V$ is the
volume of the fluid. For $n = 0$ the VCG equation of state reduces
to the original Chaplygin gas equation of state. The value of $n$
may be positive or negative. Guo \emph{et al}~\cite{guo1} showed that the
best fit value of $n = -3.4$ using \emph{gold sample of 157 SNe Ia
data}. Later in another article ~ \cite{guo2} they constrained on
VCG and determine the best fit value of $n = 0.5^{+1.0}_{-1.1}$
using \emph{gold sample of 157 SNe Ia data} and \emph{X-ray gas mass fraction in 26 galaxy cluster} ~\cite{allen}.
This result favors a
phantom-like Chaplygin gas model which allows for the possibility of
the dark energy density increasing with time. Relevant to mention
that recently there are some indications that a strongly negative
equation of state, $w \leq - 1$, may give a good fit ~\cite{beng,
novo, asta} with observations. But in  another work ~\cite{seth}, we
have seen that the value of $n$ lie in the interval $(- 1.3, 2.6)$
[\emph{WMAP 1st Peak + SNe Ia(3$\sigma$})] and $(- 0.2, 2.8)$
[\emph{WMAP 3rd Peak + SNe
Ia(3$\sigma$})].\\

Recently Santos \emph{et al}~\cite{santos} have studied the thermodynamical
stability in generalised Chaplygin gas model. In the present work we
investigate thermodynamical behaviour  of the variable Chaplygin gas
(VCG) by introducing the integrability condition equation (3) and
the temperature of equation (16). All thermal quantities are derived
as functions of  temperature and volume. In this case, we show that
the third law of thermodynamics is satisfied with the Chaplygin gas.
Furthermore, we find a new general equation of state, describing the
Chaplygin gas as function of either volume or temperature
explicitly. For the variable Chaplygin gas we expect to have similar
behaviours as the Chaplygin gas did show. Consequently, we confirm
that Chaplygin gas could show a unified picture of dark matter and
energy which cools down through the universe expansion. Returning to
the stability criterion of the Chaplygin gas we find that the value
of $n$ should be negative.   Interestingly Guo \emph{et al} ~\cite{guo1}
showed that the best fit value of $n =  -3.4$ from probability
contour. As mentioned earlier ~\cite{seth} the best fit value of $n$
may be positive or negative. From the thermodynamical stability
considerations we can constrain the value of $n$ and found that $n$
should always have negative value. The paper is organised as
follows: in section 2 we build up the thermodynamical formalism of
the VCG model and discuss the thermodynamical behaviour of this
model. Finally in section 3 we give a brief discussion.\\

\section*{2. Formalism}
\vspace{0.1 cm}
 Before proceeding  further we define the uniform density of the fluid filling the universe as

\begin{equation}
\rho = \frac{U}{V}
\end{equation}
where $U$ and $V$ are the internal energy and volume filled by the
fluid respectively. Now the energy  $U$ and pressure $P$ of
Variable Chaplygin gas may be taken as a function of its entropy $S$
and volume $V$. From general thermodynamics ~\cite{landau}, one has
the following relationship

\begin{equation}
\left(\frac{\partial U }{\partial V}\right)_{S}= - P
\end{equation}
With the help of  equations (1) - (3) we get

\begin{equation}
\left(\frac{\partial U }{\partial V}\right)_{S}=
B_{0}V^{-\frac{n}{3}} \frac{V}{U}
\end{equation}
Integrating we get

\begin{equation}
U = \left[ \frac{6B_{0}V^{\frac{6-n}{3}}} {6-n} + c
\right]^{\frac{1}{2}} = \left(
\frac{2B_{0}V^{-\frac{n}{3}}}{N}\right)^{\frac{1}{2}} V \left\{1 +
\left(\frac{\epsilon}{V} \right)^N \right\}^{\frac{1}{2}}
\end{equation}
the parameter $c$ is the integration constant which may be a
universal constant or a function of entropy $S$ only; $c = c(S)$ and
$B_{0} = B_{0}(S)$. The term $N = \frac{6 - n}{3}$ and $\epsilon =
\left[\frac{N c}{2B_{0}} \right]^{\frac{1}{N}}$ which has the
dimension of volume.  Now the energy density $\rho$ of the VCG
 reduces to the following form

\begin{equation}
\rho = V^{-\frac{n}{6}}\left[\frac{6B_{0}}{6-n} + c
V^{-\frac{6-n}{3}} \right]^{\frac{1}{2}} = \left(
\frac{2B_{0}V^{-\frac{n}{3}}}{N}\right)^{\frac{1}{2}} \left\{1 +
\left(\frac{\epsilon}{V} \right)^N \right\}^{\frac{1}{2}}
\end{equation}\\
Now we want to discuss the thermodynamical behaviour of this
model.\\

\textbf{(a) Pressure :}\\

The pressure $P$ of the VCG is also determined as a function of
entropy $S$ and volume $V$  in the following form

\begin{eqnarray}
P = - \frac{B_{0}V^{-\frac{n}{6}}}{\left[\frac{6B_{0}}{6-n} + c
V^{-\frac{6-n}{3}} \right]^{\frac{1}{2}}} = -
\left[\frac{NB_{0}V^{-\frac{n}{3}}}{2} \right]^{\frac{1}{2}}\left[ 1
+ \left(\frac{\epsilon}{V}\right)^{N}\right]^{-\frac{1}{2}}
\end{eqnarray}

\begin{figure}[ht]
\begin{center}
   \includegraphics[width=8cm]{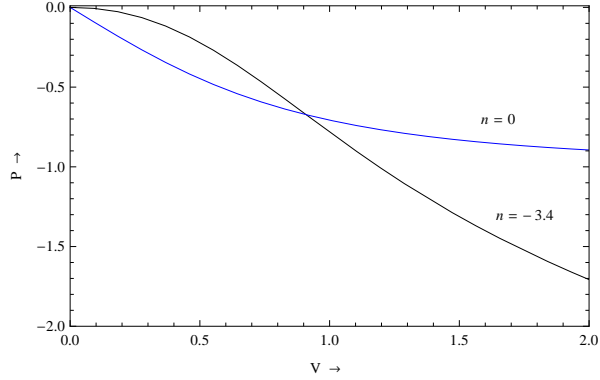}
     \caption{
  \small\emph{
  The nature of variations of $P$ and $V$  for different
  values of $n$  are shown. This figure shows that $P$ goes more and
  more negative with $V$.(Taking $B_0= 1, c = 1$).
     }\label{1}
    }
\end{center}
\end{figure}

For $n=0$ , the above results reduce to CG model ~\cite{kamen} and
its thermodynamical behaviour was discussed earlier by Santos et al
~\cite{santos}.\\

It is seen from the fig-1 that for $n \leq 0$, the pressure goes
more and more negative with volume.  We get
$P = 0$ at $V = 0$ for any value of $n$. It also follows from the
fig-1 that as $n$ becomes more and more negative
the pressure falls sharply.\\

\textbf{(b)  Equation of state:}\\

Now  from the equations (6) and (7) we  get the effective equation
of state as
\begin{equation}
\mathcal{W }= \frac{P}{\rho} = - \frac{N}{2} \frac{1}{1 +
\left(\frac{\epsilon}{V}\right)^N }
\end{equation}

\begin{figure}[ht]
\begin{center}
   \includegraphics[width=8cm]{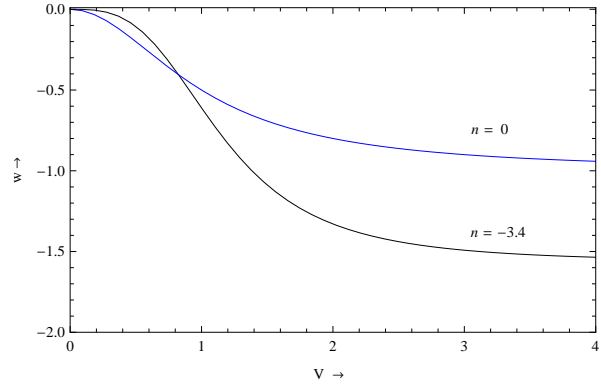}
     \caption{
  \small\emph{
  The variations of $\mathcal{W }$ and $V$  for different
  values of $n$  are shown. This figure shows that  a quiescence
  phenomenon for $n = 0$ and phantom-like phenomenon for $n <0$. (Taking $B_0= 1, c = 1$). }\label{1}
    }
\end{center}
\end{figure}

i)  For small volume, $V  \ll \epsilon$, \emph{i.e.},
$\frac{\epsilon}{V} \gg\ 1$

$\mathcal{W} \rightarrow 0 $, therefore $P \approx 0$. we get dust
dominated universe and the EoS is independent of $n$.

 ii) For large volume, $V \gg \epsilon$,
\emph{i.e.}, $\frac{\epsilon}{V} \ll 1$

\begin{equation}
\mathcal{W }\approx -1 + \frac{n}{6}
\end{equation}
 it follows from equation (9) that if  $n < 6$,  $\mathcal{W }$ is
 always greater than $ - 1$. So this is not $\Lambda$CDM, but for $n=
 0$, this will be $\Lambda$CDM. Influence of $n$ is prominent in
 this case. From equation (9) it follows that for positive values of
 $n$, the value of $\mathcal{W }$ will be $0 > \mathcal{W } > -1$. So we get a quiescence phenomenon
 and the big rip is avoided. However, in what follows we shall
 presently see that to preserve the thermodynamic stability of VCG
 $n$ should be negative. For $n <0$ we get $\mathcal{W } < -1$, the phantom-like model.
It is seen from the fig-2 that $\mathcal{W }$ is more negative for
$n = -3.4$. In an earlier work ~\cite{dp1} the present author
studied modified Chaplygin gas in higher dimensional space time and
showed that in the presence of extra dimension the model became
phantom like, but when the extra dimension is absent our results
seem to be
$\Lambda$CDM. One may see that the experimental results favour like VCG model ~\cite{guo1, guo2, seth}.\\

 \textbf{(c) Deceleration parameter:}\\

 Now we calculate the deceleration constant.
\begin{equation}
q = \frac{1}{2} + \frac{3}{2} \frac{P}{\rho} = \frac{1}{2} - \frac{3
N}{4} \frac{1}{1 + \left(\frac{\epsilon}{V}\right)^N }
\end{equation}

\begin{figure}[ht]
\begin{center}
   \includegraphics[width=8cm]{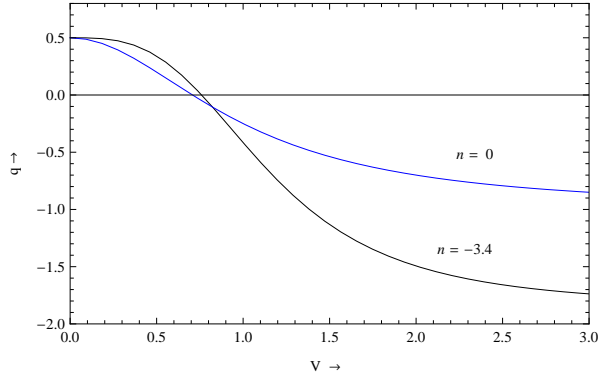}
     \caption{
  \small\emph{
  The variations of $q$ and $V$  for different
  values of $n$  are shown. Early flip is shown for $n = 0$. (Taking $B_0= 1, c = 1$).
     }\label{1}
    }
\end{center}
\end{figure}

i) For small volume, $V  \ll \epsilon$, \emph{i.e.},
$\frac{\epsilon}{V} \gg\ 1$ which gives $ q\approx  \frac{1}{2}$
\emph{i.e}., $q$ is positive, universe decelerates for small
$V$.\\

 ii) For large volume, $V \gg \epsilon$,
\emph{i.e.}, $\frac{\epsilon}{V} \ll 1$ which gives $ q\approx -1 +
\frac{n}{4}$ \\

Thus we see that initially, \emph{i.e.}, when volume is very small
there is no effect of $n$ on $q$. $q$  is positive, universe
decelerates. From fig-3 it follows that as volume increases $q$ goes
to zero first and then universe accelerates. For \emph{flip} to
occur the flip volume ($V_{f}$) is in the following form

\begin{equation}
V_{f} = \epsilon \left[\frac{2}{4-n}\right]^{\frac{1}{N}}
\end{equation}

 A little
analysis of the  equation (11)  shows that for $V_{f}$ to have real
value $n < 4$.  Otherwise there will be no \emph{flip}. This is in
accord with
the findings of the observational result~\cite{guo1}. \\

\textbf{(d) Stability:}\\

To verify the thermodynamic stability conditions of a fluid along
its evolution, it is necessary (i) to determine   if the pressure
reduces through an adiabatic expansion $ \left( \frac{\partial
P}{\partial V}\right)_{S} < 0 $ and (ii) to examine if the thermal
capacity at constant volume, $c_{V} > 0$ ~\cite{landau}. Using
equations (1) and (7) we get

\begin{eqnarray}
\left( \frac{\partial P}{\partial V}\right)_{S} =
\frac{P}{6V}\left[(6-n) \left\{1 - \frac{1}{1 +
\left(\frac{\epsilon}{V} \right)^N}\right \} -n \right]
\end{eqnarray}

\begin{figure}[ht]
\begin{center}
   \includegraphics[width=8cm]{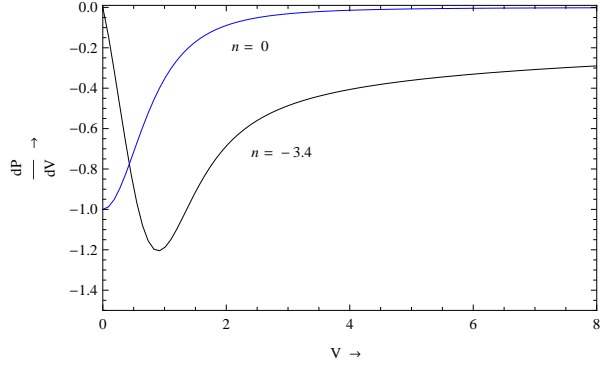}
     \caption{
  \small\emph{
  The variations of $\frac{dP}{dV}$ and $V$  for different
  values of $n$  are shown. The nature of evolution of graphs are
  quite different for $n =0$ and $n = -3.4$ but they give $\frac{dP}{dV} <0$
  throughout the evolution. (Taking $B_0= 1, c = 1$).
     }\label{1}
    }
\end{center}
\end{figure}
\vspace{0.1 cm}

In an earlier work  Sethi \emph{et al} \cite{seth} showed that the range of
$n$ lies in the interval $(- 1.3, 2.6)$ [WMAP 1st Peak + SNe
Ia(3$\sigma$)] and  $(- 0.2, 2.8)$ [WMAP 3rd Peak + SNe
Ia(3$\sigma$)]. But from equation (12) we see that for $ n \leq 0$,
$\left( \frac{\partial P}{\partial V}\right)_{S} < 0$ throughout the evolution. Fig-4 gives
similar type conclusion. So the positive value of $n$ is not
compatible in VCG model.  It may be concluded that to get
thermodynamical stable evolution the positive value of $n$ should be
discarded. One may mention that the nature of evolution of graphs
are quite different initially  for $n =0$ and $n = -3.4$ but they
both give $\frac{dP}{dV} <0$
throughout the evolution. This is due to the influence of $n$.\\

Now we should also verify if the thermal capacity at constant
volume $c_{V}$ is always positive.  First, we determine the
temperature $T$ of the Variable  Chaplygin gas as a
    function of its volume $V$ and its entropy $S$. The temperature
    $T$ of this fluid is determined from the relation  $T = \left(\frac{\partial U }{\partial S}
    \right)_{V}$. Using this relation of the
    temperature and with the help of equation (5) we get the
    expression of $T$ as follows
\begin{equation}
T = \frac{1}{2} \left[ \frac{V^{N}}{N}\frac{dB_{0}}{dS} +
\frac{dc}{dS}
     \right]\left[\frac{2B_{0} V^{N}}{N} + c
\right]^{-\frac{1}{2}}
\end{equation}

If $c$ and $B_{0}$ are also assumed to be  universal constants,
then $\frac{dc}{dS}=0$ and $\frac{dB_{0}}{dS}=0$, the fluid, in
such condition, remains at zero temperature for any value of its
volume and pressure. Therefore, to check the
thermodynamic stability of the variable Chaplygin gas whose
temperature varies during its expansion, it is necessary to assume
that the derivatives in equation (13) are not simultaneously zero.
We have no \emph{apriori} knowledge of the functional dependance of $B_{0}$ and  $c$ on $S$.
From physical considerations, however, we know that this function must
be such as to give positive temperature and cooling along an
adiabatic expansion, and we choose that $\left(\frac{\partial c
}{\partial S} \right)>0$.

Now from dimensional analysis, we observe that $[c]^\frac{1}{2} =
[U]$ which implies $[c]^{\frac{1}{2}} = [U] = [T][S]$. Thus
\begin{equation}
c = \frac{1}{\beta^2}S^2
\end{equation}
Here $\beta^{-1}$   is a universal constant with the dimension of the inverse of the
temperature: $\beta^{-1} = \tau$. Differentiating equation (14) we
get

\begin{equation}
\frac{dc}{dS} = 2 \tau^2 S
\end{equation}
In order to have positive temperatures and cooling along an
adiabatic expansion, we must impose for mathematical simplicity $\frac{dB_{0}}{dS} = 0$, which
makes the constant $B_{0}$  a universal constant. In that case equation (13) reduces to

\begin{equation}
T \approx \frac{1}{2}\frac{dc}{dS}\left(\frac{2B_{0}V^{N}}{N} + c
\right)^{- \frac{1}{2}} = \tau^{2}S \left(\frac{2B_{0}V^{N}}{N} +
\tau^2 S^2 \right)^{- \frac{1}{2}}
\end{equation}
After straight forward calculations we get the expression of
entropy as

\begin{equation}
 S =
 \left(\frac{2B_{0}}{N}\right)^{\frac{1}{2}}V^{\frac{N}{2}}\frac{T}{\tau^2}\left(1
 - \frac{T^2}{\tau^2}\right)^{-\frac{1}{2}}
\end{equation}
For positive and finite entropy $0 < T < \tau$. Evidently at $T =
0$, $S = 0$ implying that the third law of thermodynamics is
satisfied.

The thermal capacity at constant volume can be written as

\begin{equation}
c_{V} = T \left(\frac{\partial S}{\partial T} \right)_{V} =
\left(\frac{2B_{0}}{N}\right)^{\frac{1}{2}}V^{\frac{N}{2}}\frac{T}{\tau^2}\frac{1}{\left(1
 - \frac{T^2}{\tau^2}\right)^{\frac{3}{2}}}
\end{equation}

Since, $0 < T < \tau$, $c_{V} > 0$ is always satisfied irrespective of the value of $n$.\\

\textbf{ (e) Thermal equation of State:}\\

Since $P = P (T, V)$,  using equations (5), (14) and (17) we get the
internal energy as a function of both $V$ and $T$ as follows

\begin{equation}
U = \left(\frac{2 B_{0}}{N} \right)^{\frac{1}{2}}V^{ \frac{N}{2}}
\left(1 - \frac{T^2}{\tau^2} \right)^{-\frac{1}{2}}
\end{equation}
Now with the help of equations (1), (2) and (19) the pressure will
be
\begin{equation}
P = - \left(\frac{B_{0}N}{2} \right)^{\frac{1}{2}}V^{- \frac{n}{6}}
\left(1 - \frac{T^2}{\tau^2} \right)^{\frac{1}{2}}
\end{equation}
which is also a function of both $V$ and $T$.  For $T = \tau$, $P
= 0$, the universe behaves like a dust-like or a pressureless
universe, as the Chaplygin gas equation of state can not explain the primordial universe.
 Unlike the work of Santos \emph{et al} ~\cite{santos} we do not
get de Sitter like universe due to the presence of the term $
V^{-\frac{n}{6}}$ in equation (20) for the case of $ T \rightarrow
0$. Again we have seen that the isobaric curve for the VCG  do not
coincide with its isotherms  in the diagram of thermodynamic
states. Now using equation (2) and (19) we further get,

\begin{equation}
\rho = \left(\frac{2 B_{0}}{N} \right)^{\frac{1}{2}}V^{-
\frac{n}{6}} \left(1 - \frac{T^2}{\tau^2} \right)^{-\frac{1}{2}}
\end{equation}
We find exactly similar expressions of $\rho$ with the help of
equations (1) and (20). From equations (20) and (21) we get the
thermal equation of state parameter

\begin{equation}
\omega = \frac{P}{\rho}= \left( -1 + \frac{n}{6} \right) \left(1 -
\frac{T^2}{\tau^2} \right)
\end{equation}
This thermal equation of state parameter is an explicit function of
temperature only and it is also depends on $n$. As volume increases
temperature falls during adiabatic expansions. In our case, for $ T
\rightarrow 0$, the equation (22) yields $ \omega = - 1 +
\frac{n}{6}$ which is identical with the equation (9) as it is the
case of large volume. Again as $ T \rightarrow \tau$ (the maximum temperature), $\omega
\rightarrow 0$ which is indicating dust dominated universe as expected.

Now we have to examine the well known  thermodynamical relation  as

\begin{equation}
\left(\frac{\partial U}{\partial V}\right)_{T} = T
\left(\frac{\partial P}{\partial T}\right)_{V} -P
\end{equation}
Using equations (19) and (20) we find the relation (23) is also satisfied.\\

We can also express the maximum temperature $\tau$ as a function of the initial conditions of the expansion.
If we consider that the initial conditions at $V = V_0$ are $\rho = \rho_{0}$, $P = P_{0}$ and $ T = T_{0}$, then we can get from equation (5) as

\begin{equation}
c = \left(\rho_{0}^2 - \frac{2 B_{0}}{N} V_{0}^{-\frac{n}{3}} \right) V_{0}^2
\end{equation}
With the help of equations (6), (7) and (24), we obtained the energy density $\rho$  and the pressure $P$ as a function of the volume $V$ as
\begin{equation}
\rho = V^{- \frac{n}{6}} \rho_{0}\left[\frac{2B_0}{N \rho_{0}^2} + \left( 1- \frac{2B_0}{N \rho_{0}^2}  V^{- \frac{n}{3}} \right) \left( \frac{V_{0}}{V} \right)^2  V^{ \frac{n}{3}} \right]^{\frac{1}{2}}
\end{equation}

and
\begin{equation}
P =  - \frac{B_{0}^{\frac{1}{2}}\left(\frac{B_{o}}{\rho_{0}^2} \right)^{\frac{1}{2}} V^{- \frac{n}{6}}} {\left[\frac{2B_0}{N \rho_{0}^2} + \left( 1- \frac{2B_0}{N \rho_{0}^2}  V^{- \frac{n}{3}} \right) \left( \frac{V_{0}}{V} \right)^2  V^{ \frac{n}{3}} \right]^{\frac{1}{2}}}
\end{equation}
Now the equations (20), (25) and (26) can be written as function of the reduced parameters $\varepsilon$,  $v$,  $p$, $ \kappa$, and $t$ such that

\begin{eqnarray}  \nonumber
\varepsilon = \frac{\rho}{\rho_{0}} , ~~~~~ v = \frac{V}{V_{0}}, ~~~~~ p= \frac{P}{B_{0}^{\frac{1}{2}}},\\
\kappa = \frac{2 B_{0}}{N \rho_{0}^2} , ~~~~~ t = \frac{T}{T_{0}}, ~~~~~  \tau^{*} = \frac{\tau}{T_{0}}
\end{eqnarray}
\vspace{0.5cm}
the equations (20), (25) and (26)  can be written in the reduced units  respectively as

\begin{equation}
p = -\left(\frac{N}{2} \right)^{\frac{1}{2}} V^{-\frac{n}{6}} \left(1 - \frac{t^2}{\tau^{*2} }\right)^{\frac{1}{2}}
\end{equation}

\begin{equation}
\varepsilon = V^{-\frac{n}{6}} \left[ \kappa + \left( 1 - \kappa  V^{-\frac{n}{3}} \right) \frac{ V^{\frac{n}{3}}}{v^2} \right]^{\frac{1}{2}}
\end{equation}

\begin{equation}
p = - \frac{\kappa^{\frac{1}{2}} \left(\frac{N}{2} V^{-\frac{n}{3}} \right)^{\frac{1}{2}}}{\left[ \kappa + \left( 1 - \kappa  V^{-\frac{n}{3}} \right) \frac{ V^{\frac{n}{3}}}{v^2} \right]^{\frac{1}{2}}}
\end{equation}

At $P = P_{0}$,  $V = V_{0}$ and  $T = T_{0}$, we have $ t = 1$ and $v = 1$, and we get from equations (28) and (30)

\begin{equation}
p_{0} = - \kappa^{\frac{1}{2}} \left(\frac{N}{2}\right) ^{\frac{1}{2}}V_{0}^{-\frac{n}{3}}  = - \left(\frac{N}{2}\right) ^{\frac{1}{2}} V_{0}^{-\frac{n}{6}} \left( 1 - \frac{1}{\tau^{*2}} \right)^{\frac{1}{2}}
\end{equation}
hence  $\kappa$ and $\tau^{*}$ can be determine as follows

\begin{equation}
\kappa = V_{0}^{\frac{n}{3}} \left( 1 - \frac{1}{\tau^{*2}} \right)
\end{equation}
and
\begin{equation}
\tau^{*} = \frac{1}{\left( 1- \kappa  V_{0}^{- \frac{n}{3}}\right)}
\end{equation}
Interestingly, we have seen that $\tau^{*}$  is depend on both $\kappa$, $V$ and $n$ also. For $ n = 0$, all the above equations reduce to the equations of Santos \emph{et al} ~ \cite{santos}. At present epoch, $\kappa = \frac{2 B_{0}}{N \rho_{0}^2}$, therefore, $\rho_{0} = \left(\frac{2 B_{0}}{N \kappa} \right)^{\frac{1}{2}}$.  If we considered that the temperature $\tau = 10^{32}$K (temperature of Planck era) and $T_{0} = 2.7 $K (the temperature of the present epoch), the ratio, $\tau^{*} = \frac{\tau}{T_{0}} = 3.7 \times 10^{31} $. So the ratio $\kappa $ will be,

\begin{equation}
\kappa = V_{0}^{\frac{n}{3}} \left[ 1 - \frac{1}{\left( 3.7 \times 10^{31} \right)^{2}} \right] \approx V_{0}^{\frac{n}{3}}
\end{equation}

Again from equation (21), for the case of present epoch when temperature $T$ is very small ( i.e., $T \rightarrow 0$),
\begin{equation}
\rho_{0} \approx \left(\frac{2 B_{0}}{N V_{0}^{\frac{n}{3}} } \right)^{\frac{1}{2}} \approx \left(\frac{2 B_{0}}{N \kappa } \right)^{\frac{1}{2}}
\end{equation}
The same result can be obtained from equation (6) for large volume.

Thus, consideration from equation (14), at the present epoch, the energy density $\rho$ of the universe filled with the VCG must be very close to $\left(\frac{2 B_{0}}{N \kappa } \right)^{\frac{1}{2}}$.
\section*{3. Discussion}
\vspace{0.3cm}
 We have studied thermodynamical behaviour of VCG model. We consider
 the value of $ n = - 3.4$ which was found  by Guo \emph{et al}~ \cite{guo1}. In an
 earlier literature we have seen that the value of $n$ may be
 negative or positive ~\cite{seth}. For large volume, when $n > 0$ the effective equation of state
 results in a quiescence type whereas $n = 0$ goes to $\Lambda$CDM model.
 Again for $n <0$ it favours phantom-like cosmology. Some important
 results are given below:\\

 i) As we have considered $n = -3.4$, the pressure goes more and more
 negative as volume increases (fig-1). \\

 ii) The effective equation of state is shown in equation (8). At
 large volume we have seen that  for $n = 0$, gives $\Lambda$CDM
 model.  Influence of $n$ is prominent for sufficiently large
 volume. For $n < 0$, $\mathcal{W} <0$ leads to phantom like
 cosmology which is in favour of the current SNe Ia constraints like
 VCG model. The above phenomena is shown in fig-2.\\

 (iii)  The deceleration parameter is studied in the context of
 thermodynamics and is shown in fig-3. Our analysis shows that for the \emph{flip} to occur
  the value of $ n < 4$. This is in accord with the
 findings of the observational results ~\cite{guo1}.\\

 (iv) The most important area of our concern is the question of the
 thermodynamic stability of the gas chosen.  Firstly, we have to
 determine whether $ \left(
\frac{\partial P}{\partial V}\right)_{S} < 0 $. The analysis shows
that  only for the negative value of $n$, $ \left( \frac{\partial
P}{\partial V}\right)_{S} < 0 $ throughout the evolution. So one important conclusion done
here using equation (12) is that
 the value of $n$ must be negative. Interestingly, this result is in
 agreement with the observational result found earlier by Guo \emph{et al}~\cite{guo1}.
  Due to this
 reason we have taken $n = -3.4$ to study the whole work done
 in this article. In this context the thermal capacity at constant
 volume $c_{V}$
 is also determined and it is seen that $c_{V}$ is always positive irrespective of the value of
 $n$. So both the conditions of thermodynamic stability of the fluid are studied
 which shows that the fluid is thermodynamically stable through out the evolution process.\\

 (v) The expression of entropy is derived and shown in equation
 (17). In this equation it is seen that at $T =
0$, $S = 0$ implying that the third law of thermodynamics is
satisfied.\\

 (vi) Finally the  thermal equation of state is discussed in this work  where
 it is seen that the volume is not explicitly  present in the
 expression (22).  This thermal equation of state  parameter
 is an explicit function of temperature
only. As volume increases temperature falls during adiabatic
expansions. In this case, for $ T \rightarrow 0$, the equation
(22) yields $  \omega = - 1 + \frac{n}{6} $ which is identical
with the equation (9) as it is the case of large volume. Again as
$ T \rightarrow \tau$, $\omega \rightarrow 0$ pointing to a dust
dominated universe. As this type of equation of state can not explain the primordial universe.
Here the maximum temperature $\tau$ is expressed as a function of the initial conditions of the expression.

\vspace{0.5cm}

\textbf{Acknowledgment : }  DP acknowledges the financial support of
UGC, ERO for a MRP ( No- F-PSW- 165/13-14) and also  CERN, Geneva, Switzerland
for a short visit. The author thanks   the referee  for valuable comments and suggestions.


\begin{thebibliography}{40}
\bibitem{reis} A G Reiss et al,  \emph{Astron. J.} \textbf{607} 665
(2004)[astro-ph/9805201].

\bibitem{aman} R Amanullah et al,  \emph{Astrophy. J.} \textbf{716}
712 (2010).

\bibitem{kamen} A Kamenschick, U Moschella, V. Pasquier, \emph{Phys. Lett.} \textbf{B511} 265 (2001).


\bibitem{fab} J C Fabris, S V B Goncalves, P E de Souza,  \emph{Gen. Relativ, Grav.} \textbf{34} 53 (2002).

\bibitem{bena} H. B. Benaoum, (2002) [hep-th/ 0205140].

\bibitem{lima} J. A. S. Lima, J. V. Cunha and J. S. Alcaniz, (2006) [astro-ph
/0608469v1].

\bibitem{as} M. C. Bento, O. Bertolami and A. A. Sen  Phys.
Rev. D\textbf{66} 043507 (2002).

\bibitem{dp2} D. Panigrahi and S. Chatterjee,  JCAP \textbf{10} 002 (2011);

\bibitem{guo1} Zong-Kuan Guo  and Yuan-Zhong Zhang, \emph{Phys. Lett.} \textbf{B645} 326 (2007).

 \bibitem{guo2} Zong-Kuan Guo  and Yuan-Zhong Zhang, (2005) [astro-ph/0509790 ].

 \bibitem{allen} S W Allen et al, Mon. Not. Roy.Astron. Soc. \textbf{353} 457 (2004).

 \bibitem{beng} G R Bengochea, \emph{Phys. Lett.} \textbf{B695} 405 (2011).


 \bibitem{novo} B Novosyadlyj, O Sergijenko, R Durrer, V Pelykh, \emph{Phys. Rev.} \textbf{D86} 083008 (2012).

 \bibitem{asta} A V Astashenok, S Nojiri, S D Odintsov, A V Yorov, \emph{Phys. Lett.} \textbf{B709} 396 (2012).


\bibitem{seth} Geetanjali Sethi, Sushil K. Singh, Pranav Kumar \emph{Int. Jour. Mod. Phys.}\textbf{D15} 1089
(2006); [astro-ph/0508491].


\bibitem{santos} F C Santos, M L Bedran, V Soares \emph{Phys. Lett.} \textbf{B636} 86
 (2006).

\bibitem{landau} L  D Landu, E  M  Lifschitz, statistical Physics, third ed.,
 Course of Theoretical Physics, vol. 5, Butterworth-Heinemann, London, 1984.


 \bibitem{dp1} D Panigrahi and S Chatterjee \emph{Grav. Cosml.} \textbf{17} 81
 (2011) [grqc /1006.0476]; D Panigrahi , Proceedings No: CP  1316, Search
 for fundamental theory, edited by R L Amoroso, P Rowlands and S
 Jeffers , AIP, 2010.











\end{thebibliography}
\end{document}